# Hydrostatic pressure effect on structural and transport properties of co-existing layered and disordered rock-salt phase of $Li_xCoO_2$

Thiagarajan Maran[1], A. Jain[2,3] *, Muthukumaran Sundaramoorthy[1,5], A. P. Roy [4], Boby Joseph[5], Govindaraj Lingannan[1], Ashwin Mohan [6], D. Bansal[4], S. M. Yusuf[2,3] †, and Arumugam Sonachalam [1,7] ‡

[1]Center for High Pressure Research, Bharathidasan University, Tiruchirappalli - 620 024, India

[2]Solid State Physics Division, Bhabha Atomic Research Centre, Mumbai 400 085, India

[3]Homi Bhabha National Institute, Anushaktinagar, Mumbai 400 094, India

[4]Department of Mechanical Engineering, Indian Institute of Technology Bombay, Mumbai 400 076, India

[5] Elettra - Sincrotrone Trieste S.C. p.A., S.S. 14, Km 163.5 in Area Science Park, Bosavizza 34149, Italy

[6]Department of Physics, Institute of Chemical Technology, Matunga, Mumbai 400019, India.

[7]Tamil Nadu Open University, Chennai- 600 015, India

*ajain@barc.gov.in, †smyusuf@barc.gov.in, ‡ sarumugam1963@yahoo.com

*It is widely believed that the origin of a significant cause for the voltage and capacity fading observed in lithium (Li)-ion batteries is related to structural modifications occurring in the cathode material during the Li-ion insertion/de-insertion process. The Li-ion insertion/de-insertion mechanism and the resulting structural changes are known to exert a severe strain on the lattice, and consequently leading to performance degradation. Here, with a view to shed more light on the effect of such strain on the structural properties of the cathode material, we have systematically investigated the pressure dependence of structural and transport properties of an $Li_xCoO_2$ single crystal, grown using 5% excess Li in the precursors. Ambient pressure synchrotron diffraction on these crystals reveals that, the excess Li during the growth, has facilitated the stabilization of a layered rhombohedral phase (space group $R\bar{3}m$) as well as a*

*disordered rock-salt phase (space group Fm$\bar{3}$m). The volume fraction of the rhombohedral and cubic phase is 60:40, respectively, which remains unchanged up to 10.6 GPa. No structural phase transition has been observed up to 10.6 GPa. An increase in resistance with a decrease in temperature has revealed the semi-metallic nature of the sample. Further, the application of hydrostatic pressure up to 2.8 GPa shows the enhancement of semi-metallic nature. The obtained experimental results can be qualitatively explained via density functional theory (DFT) and thermodynamics modelling. The calculated density of states was reduced, and the activation energy was increased by applied pressure. Our investigations indicate a significant phase stability of the mixed phase crystals under externally applied high pressure and thus suggest the possible use of such mixed phase materials as a cathode in lithium-ion batteries.*

# 1. INTRODUCTION:

LiCoO$_2$ has been extensively used as a typical cathode material in lithium-ion batteries due to a high theoretical energy density and significant structural stability during insertion and removal of lithium ions [1,2]. However, severe voltage decay of lithium-ion batteries, upon extended cycling, is one of the limiting factors regarding its utilization [3]. Therefore, tremendous efforts have been undertaken in recent years to understand the origin of the voltage fade. The delocalization of electronic states was initially suggested as a possible key factor for the voltage fade [4]. It was widely believed that Li extraction occurs via a solid solution reaction and structural transitions might occur during the process, impacting the electrical properties and mobility of lithium ions [4-6]. An insulator-to-metal-like transition was reported with a minor change in the lithium concentration [6]. However, a recent study, combining in-situ x-ray diffraction and ex-situ STEM measurements [7], indicates that a significant cause of this voltage fade is related to structural factors, specifically to the transformation from layered rhombohedral structure (space group $R\bar{3}m$) of LiCoO$_2$ to spinel-like trigonal ($P\bar{3}m1$) structure of the completely delithiated end-member CoO$_2$. The layered crystal structure of Li$_x$CoO$_2$ contains edge-sharing CoO$_6$ octahedra, separated by octahedrally coordinated Li along [001] direction. The layered structure [Fig. 1 (a)] adopts -ABCABC- (O3 phase) type stacking sequence of oxygen ions along [001] direction [7]. Upon delithiation, a collective and quasi-continuous glide of CoO$_6$ slabs was reported to trigger a transition from O3 to O1 (having -AA- type stacking sequence of oxygen) phase, before the layered to spinel-like trigonal ($P\bar{3}m1$) phase transformation. During the transition from O3 to O1, gradual angle ($\delta$), defined as the angle between the cobalt ions along [001] crystallographic axis [Fig. 1(a)], decreases to zero, from $\delta \sim 9.7°$ for O3 phase. Moreover, during the phase transformation from the O3 to O1 phase, numerous intermediate phases, including the pseudomorphic phase of Li-Co antisites (due to the gradual migration of Co ions from the CoO$_6$ slab to the Li sites) were reported to exist in local regions, especially at high delithiation levels [7]. With an increase in the Li-Co antisites, a layered-to-trigonal type transition took place, in which layered $Li_xCoO_2$ structure transformed in a trigonal phase ($P\bar{3}m1$), which was reported to have a profound influence on the capacity fading and optimization of high-energy density for lithium-ion batteries [7]. The overall phase transition was explained in terms of a gradual angle. The reported value of gradual angle was 9.712° for $x = 1$, which dropped to zero for $x = 0.3$. In-situ x-ray diffraction results reported in ref. [7] corresponds to a quasi-thermodynamic condition on powder samples, therefore obtaining quantitative information about the phase fraction of different phases becomes difficult. Due to this complexity, thus far, conclusive evidence of the structural and electronic structure details, as well as about thermodynamic phase stability of phases with different lithium content has been lacking.

In order to gain insight into these strain-induced pseudomorphic phases occurring during lithiation and de-lithiation, we investigated the effect of external pressure on the structural and transport properties of an $Li_xCoO_2$ single crystal with a view to explore the possibility of pressure-driven stabilization of such phases. The $LiCoO_2$ single crystal which was synthesized by optical zone floating technique [8], with 5% excess lithium, was subjected to high pressure to study and analyse its structural properties and electronic transport. The present study shows the co-existence of $Li_{0.12}Co_{0.88}O$ disordered rock-salt phase (space group $Fm\bar{3}m$), along with the main $LiCoO_2$ rhombohedral phase ($R\bar{3}m$) in the crushed single crystal sample. We observe that the gradual angle, derived from the Rietveld refinement analysis of the synchrotron diffraction data, is 9.848° at 0 GPa for the rhombohedral phase and increases to 9.994° at 10.6 GPa. An increase in resistance with decreasing temperature reveals the semi-metallic nature of the sample. The density of states $N(E_F)$ reduces and the activation energy ($E_a$) increases, and the derived value of piezo-resistivity increases with applied pressure. The obtained results can be qualitatively explained via density functional theory (DFT) and thermodynamics modelling. The observed increase in the gradual angle with an increase in the pressure for the rhombohedral phase, while having a constant lithium content, also suggests a novel path where one could apply internal negative pressure by using the dopants with less ion radius at Co or Li sites to synthesize $Li_xCoO_2$, for mitigating the gradual structural transition from the O3 to the O1 phase during Li insertion/de-insertion. We point out that the increase in the gradual angle must be related to the presence of intergrowth of cubic phase's crystallographic domains in the material, thereby inducing the stabilization of the rhombohedral phase in the presence of large lattice strain due to applied pressure. We find that the relative changes in lattice parameters and volume, as a function of pressure, are small for the disordered rock-salt phase ($Fm\bar{3}m$), compared to the layered rhombohedral phase ($R\bar{3}m$). Our results therefore suggest the possibility of the disordered rock-salt phase $Li_{1-x}Co_xO$ acting as a transitional phase during the Li de-insertion/insertion process to and from the layered phase and provides a rationale for the feasibility of designing such mixed phase materials for use as a cathode in lithium-ion batteries.

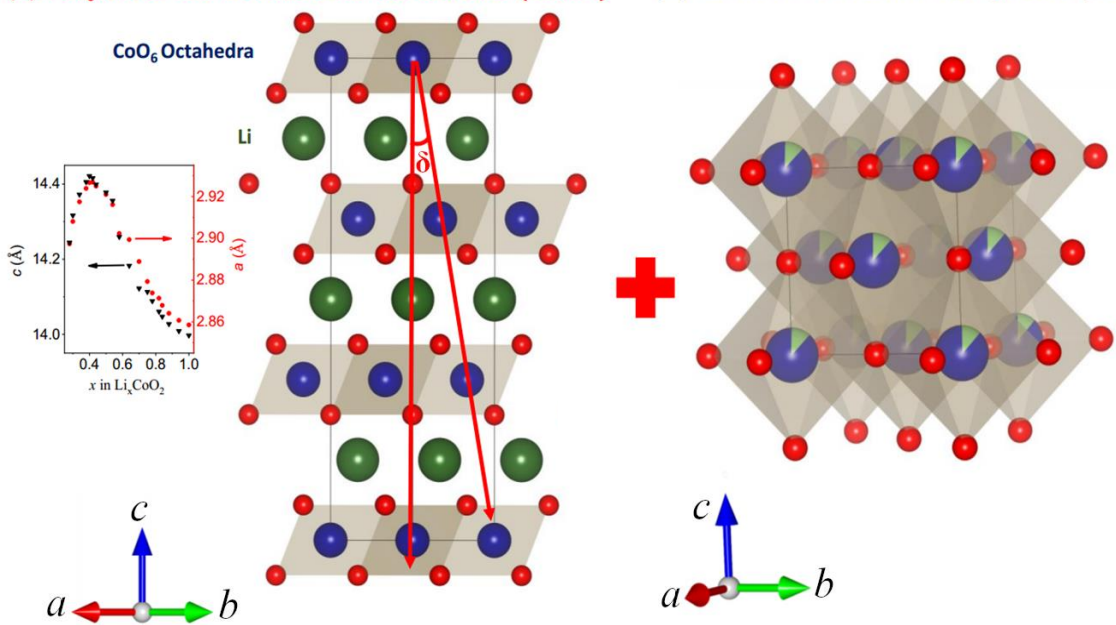

**(a) Layered rhombohedral structure ($R\bar{3}m$)**   **(b) Disordered rock-salt phase ($Fm\bar{3}m$)**

*Fig. 1 (a) Layered crystal structure of LiCoO$_2$ having -ABCABC-type (O3 phase) staking sequence along [001] axis and schematic of gradual angle (δ), inset: variation of lattice parameters with Li concentration in Li$_x$CoO$_2$ derived from structural parameters reported in ref. [7] (b) Cubic Li$_{0.12}$Co$_{0.88}$O structure, where Li and Co share the same octahedral crystallographic site.*

## 2. EXPERIMENTAL:

### 2.1. Sample Synthesis:

A conventional solid-state reaction method was used to create a polycrystalline sample of LiCoO$_2$ from 99.99% Co$_3$O$_4$ (Alfa Aesar Inc.) and 99.99% Li$_2$CO$_3$ (LTS Research Labs Inc.) as precursors. Further, it was used to prepare feed and seed rods to grow a single crystal sample. The high-quality, LiCoO$_2$ single crystals are grown in an Ar-atmosphere of 9 bar and using slow growth speeds of 1–2 mm/h, using feed rods containing 5% excess Li content by optical zone floating technique (Crystal Systems Co., Japan, FZ-T-10000-H-VII-VPO-PC) [8]. In Ref. [8] detailed techniques for synthesis were explained. The lithium composition of grown crystals was determined using the iCAP 6300 Duo inductively coupled plasma-optical atomic emission spectrometer (ICP-AES).

### 2.2. High Pressure XRD:

Ambient and high-pressure X-ray powder diffraction (HP-XRD) data at room temperature were collected at the Elettra Synchrotron Center (Xpress beamline), Trieste, Italy. A membrane-driven symmetric diamond anvil cell (DAC) with a culet size of 400 μm and a PACE-5000-based automatic membrane drive was used. The sample chamber for the HP-XRD was prepared by intending a 200 μm thickness 301 stainless steel metal foil to 50 μm and drilling a 160 μm hole in the center. A methanol-ethanol mixture with a ratio of 4:1 is used as the pressure-transmitting medium (PTM) for the experiment to ensure hydrostaticity throughout the experiment [9]. *In-situ* Ruby fluorescence technique was used to monitor the pressure inside the DAC by including a ruby ball of size ~5 μm inside the sample chamber along with the sample. A monochromatic circular beam with a wavelength of 0.4957 Å and a cross-sectional diameter of 40 μm was used to illuminate the sample along with the ruby ball. A PILATUS3S-6M large area detector was used for diffraction data collection. FIT2D software was employed for integrating the data into 2 Theta vs intensity plots. The Rietveld refinement using the GSAS-II suite was carried out to obtain the structural parameters as a function of pressure [10].

### 2.3. High Pressure Resistivity:

Ambient and high-pressure electrical resistivity measurements were done using a cryogen-free closed-cycle refrigerator (CCR-VTI) system at temperatures ranging from 4 to 300 K using a standard linear four-probe DC resistivity technique. The electrical resistivity of a rectangular bar-shaped single crystal sample with the dimensions of 1x0.75x0.01 mm$^3$. The electrical contacts were made using a high-quality silver paste and Cu wire ($\phi$ 40 μm). A clamp-type hybrid double-cylinder (NiCrAl-inner cylinder; BeCu-outer cylinder) piston pressure cell was used to generate a hydrostatic pressure. The hydrostaticity of the pressure was ensured by using Daphne oil 7474 as the pressure-transmitting medium [11]. The actual pressure that can be accumulated inside the pressure cell was given by the previously generated curve using the fixed pressure points of bismuth calibration [12]. A 20-ton hydraulic press (Riken Kiki, Japan) was used to apply the load on the pressure cell.

### 2.4. Dc Magnetization Measurements

The DC magnetization measurements were carried out in a CRYOGENIC make vibrating sample magnetometer (VSM) with the temperature ranging from 5 to 300 K and applied a magnetic field of 1000 Oe.

### 2.5. First Principles Simulations

Electronic structure simulations were performed in the framework of density functional theory (DFT) as implemented in the Vienna ab initio Simulation Package (VASP) [13-15]. The projector-augmented-wave potentials explicitly included 1 valence electron for Li ($2s^1 2p^0$), 9 for Co ($3d^8 4s^1$) and 6 for O ($2s^2 2p^4$) within generalized gradient approximation (GGA) in the Perdew-Burke-Ernzerhof (PBE) parametrization [16] with a Hubbard correction. To treat the localized $d$−electron states of Co in GGA+$U$ calculations, the total energy expression was described as introduced by Dudarev *et al.* [17] with on-site Coulomb interaction $U$. We find our calculated band gap with $U$ = 1 and 3 eV to be consistent with the experimental gap ($E_g \sim$ 2.5 [18] and 2.6 [19] eV for LiCoO$_2$ and CoO, respectively). We used a converged plane-wave cutoff energy $E_{Gcut}$ = 650 eV on a 4-atom primitive cell with a 6x6x6 and 8x8x8 Γ-centered Monkhorst-Pack electronic $k$-point mesh for LiCoO$_2$ and CoO (spin-polarized calculation), respectively. The total energy converged to less than 1 meV/atom for both $E_{Gcut}$ and $k$-mesh. The convergence criteria for a self-consistent electronic loop were set to $10^{-8}$ eV. Our results show a minimal influence of spin-orbit coupling on the ground state properties of CoO, including lattice parameters and bulk modulus. Consequently, we did not include spin-orbit coupling in our simulations, except for initial testing.

### 2.6. Maximally-Localized Wannier Functions (MLWFs):

We used WANNIER90 (v3.1.0) [20] to calculate MLWFs on a 22 (ten $d$ orbitals of two Co atoms and six $p$ orbitals of two O atoms) and 16 orbital basis (five $d$ and three $p$ orbitals of two Co and O atoms) for LiCoO$_2$ and CoO, respectively. Electron band structure using MLWFs is consistent with the direct DFT calculations. Using MLWFs, we solve the Boltzmann transport equation with BoltzWann [21] code within constant relaxation time approximation (RTA) on a converged k-mesh of 50x50x50. Electron-electron scattering time (τ) was kept constant at 10 fs.

### 3. RESULTS:
### 3.1. High Pressure XRD:

Figure 2 shows the measured powder XRD data and Rietveld refinement for LiCoO$_2$ at 0 GPa using the GSAS II suite [10]. The synchrotron pattern reveals that the LiCoO$_2$ material exhibits a rhombohedral phase with space group $R\bar{3}m$ with lattice constants $a = b = 2.815(7)$ Å, and $c = 14.057(6)$ Å and the unit cell volume of 96.51 Å$^3$. The obtained parameters are in good agreement with the reported values [22]. The co-existence of another phase apart from the rhombohedral phase has also been found. A cubic phase of space group $Fm\bar{3}m$ with the lattice constant value of $a = b = c = 4.220(7)$ Å and unit cell volume of 75.192 Å$^3$ has been extracted from the refinement.

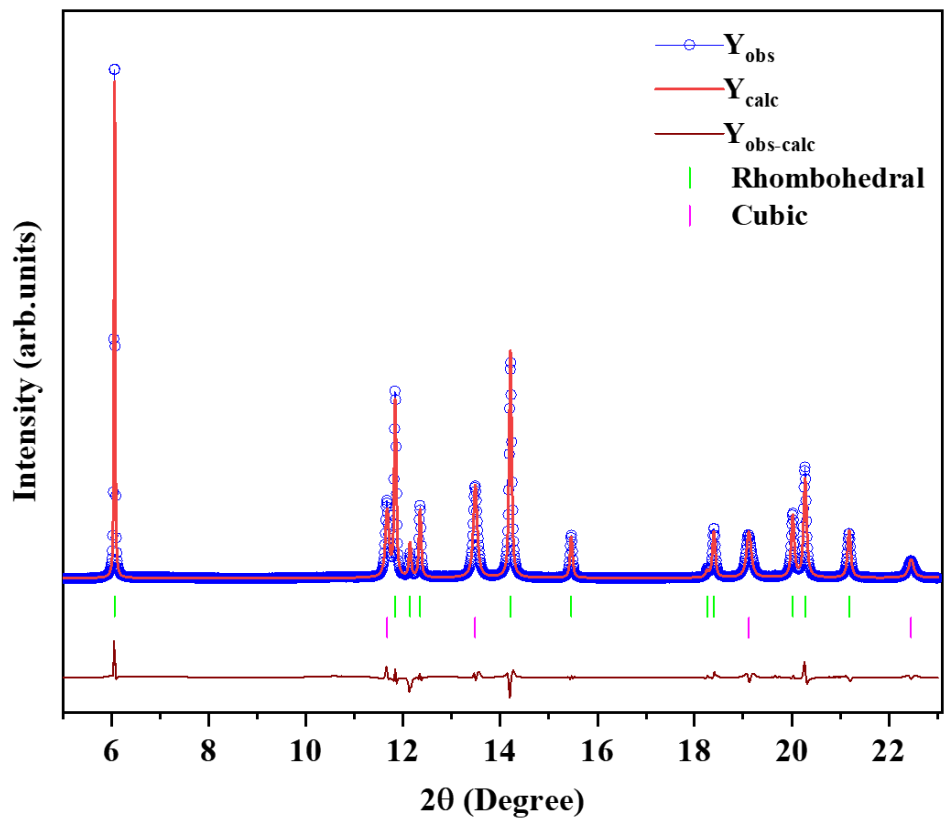

*Fig. 2 Observed (open circles) and Rietveld refined (solid line) x-ray diffraction patterns of Li$_x$CoO$_2$ at 0 GPa. Vertical bars below the pattern represent the peak positions of the two phases.*

Figure 3 shows the measured high-pressure powder XRD patterns using a synchrotron source up to 10.6 GPa which are stacked vertically for clear representation. A shift in the position of

(003) Bragg peak (2θ ~ 6.25°) to higher angles, with an increase in the applied pressure, implies a decrease of *d*-spacing. Peak shift towards the higher angle region throughout the entire process highlights that the effect of pressure remains the same up to the maximum applied pressure and shows the uniform compressibility of the sample. We did not observe the emergence or disappearance of existing Bragg peaks indicating the lack of structural phase transition up to 10.6 GPa. The result concurs with the reported results of Wang *et al.* [23] and Hu *et al.* [24]. Based on the results of the Rietveld refinement, the volume fraction of rhombohedral and disordered rocks-salt cubic phase ($Fm\bar{3}m$) is 60:40, which remains unchanged up to 10.6 GPa. The Rietveld refinement of the diffraction data, ICP-AES measurements, and dc magnetization (discussed later) confirmed that the chemical compositions of rhombohedral and cubic phase are $LiCoO_2$ and $Li_{0.12}Co_{0.88}O$, respectively. Also, the lattice parameter of 4.22 Å obtained from the refinement procedure points to the existence of $Li_{0.12}Co_{0.88}O$, upon an assessment of the experimentally derived plot of the lattice parameter against x (in $Li_{1-x}Co_xO$) [34]. Additionally, Li content obtained from ICP-OES results, can be understood as distributed between the $LiCoO_2$ rhombohedral phase and the $Li_{0.12}Co_{0.88}O$ phase. The Rietveld refinement further shows that in the cubic phase, both lithium and cobalt share the same octahedral site (Wyckoff position 6*a*, space group # 225).

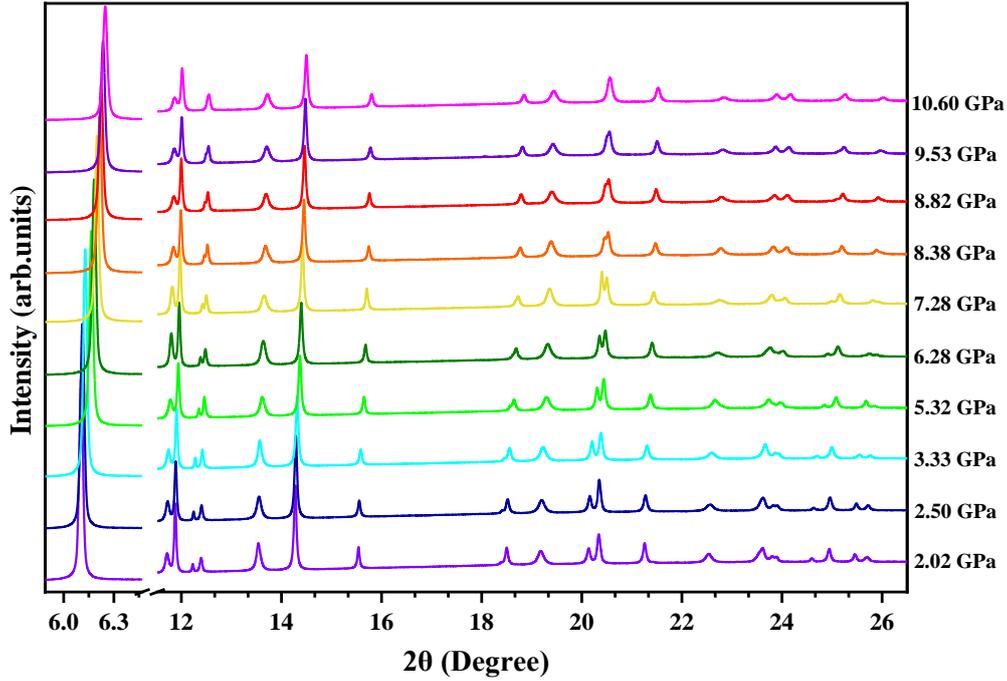

*Fig. 3* Collected synchrotron high pressure powder x-ray diffraction spectra of the sample $Li_xCoO_2$ up to 10.6 GPa using diamond anvil cell.

Figure 4(a) shows the refined lattice parameters from our XRD measurements as a function of pressure for the rhombohedral phase ($R\bar{3}m$) highlighting a linear decrease. Thus, both *a* and *c*-axes lattice parameters undergo a usual contraction upon applying pressure up to 10.8 GPa. Importantly, δ, the gradual angle for the rhombohedral phase, (see Fig. 4(b)) increases linearly from 9.848º at ambient pressure up to 9.994º at 10.6 GPa, in contrast to its gradual decrease while transiting from layered-to-rock-salt-type ($R\bar{3}m$ to $P\bar{3}m1$) while charging-discharging [7]. Similarly, upon tracking lattice parameters for the cubic phase ($Fm\bar{3}m$, see Fig. 4(d)), we find a linear decrease upon applying pressure. The trend in lattice parameters is reproduced using our simulations for the rhombohedral phase. Table 1 presents a detailed report of lattice parameters (near ambient pressures) and corresponding pressure derivatives for both rhombohedral and cubic phases. Our measured and calculated lattice parameters are consistent with earlier investigations on $LiCoO_2$ [24-26].

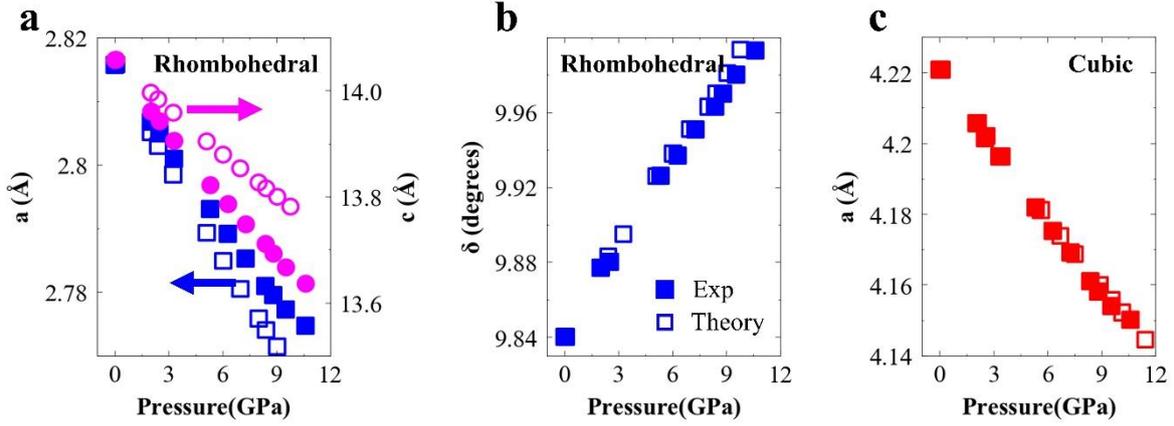

*Fig. 4 (a) a-axes (blue squares), c-axes (magenta circles) lattice parameters and (b) gradual angle (δ as discussed in the text) as a function of pressure for rhombohedral phase ($R\bar{3}m$) of LiCoO$_2$. (c) a-axis lattice parameter (red squares) for the cubic phase ($Fm\bar{3}m$). Filled and hollow markers represent measured and calculated data, respectively. A constant offset of 3 GPa is applied on calculated pressures.*

Figure 5 shows unit cell volume as a function of the applied pressure. Measured and calculated results display that the unit cell volumes shrink upon applying pressure for both rhombohedral ($R\bar{3}m$) and cubic ($Fm\bar{3}m$) phases. We fit the pressure dependent unit cell volumes with third order Birch-Murnaghan (B-M) equation of state, expressed as [27]

$$P(V) = \frac{3B_0}{2}\left[\left(\frac{V_0}{V}\right)^{\frac{7}{3}} - \left(\frac{V_0}{V}\right)^{\frac{5}{3}}\right]\left\{1 + \frac{3}{4}(B_0' - 4)\left[\left(\frac{V_0}{V}\right)^{\frac{2}{3}} - 1\right]\right\}, \quad (1)$$

where $B_0$ is the bulk modulus and $B_0'$ is the pressure derivative of $B_0$ [24-26]. B-M equation shows a good fit to the measured and calculated data, as shown in Fig. 5. Thereafter, from the obtained B-M fits, we report a detailed comparison of $B_0$ and $B_0'$ in Table 1 for both the phases. The results highlight a reasonable agreement of $B_0$ for both rhombohedral and cubic phases from our XRD measurements (141 and 171 GPa, respectively) and first principle simulations (155 and 196 GPa, respectively) with the reported values [24].

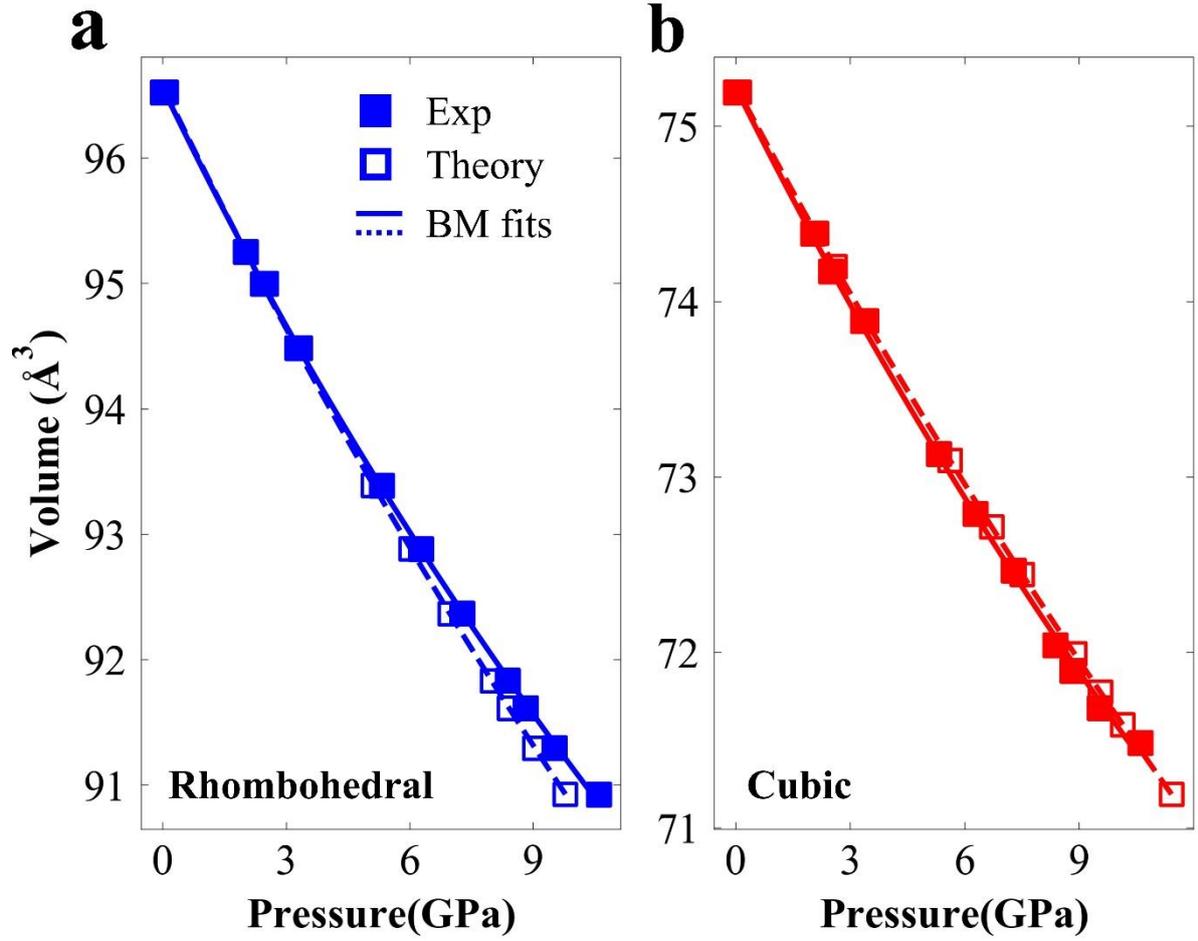

*Fig. 5* Volume as a function of pressure for (a) rhombohedral (R$\bar{3}$m, blue) and (b) cubic (Fm$\bar{3}$m, red) phases. Filled and hollow markers represent measured and calculated data, respectively. While solid and dotted lines denote the corresponding BM fits.

*Table 1*. Comparison of measured lattice parameters, volume, Bulk modulus, and corresponding pressure derivatives with calculated values for both rhombohedral and cubic phases. *a* and *c* values in simulations were taken from experimental cif file.

| Parameters | Rhombohedral phase (R$\bar{3}$m) | | Cubic phase (Fm$\bar{3}$m) | |
| --- | --- | --- | --- | --- |
| | Exp | Theory | Exp | Theory |
| *a* (Å) | 2.815 | 2.815 | 4.221 | 4.221 |
| *c* (Å) | 14.058 | 14.058 | -- | -- |
| *v* (Å³) | 96.52 | 96.52 | 75.19 | 75.19 |
| $B_0$ (GPa) | 141 | 155 | 171 | 196 |
| $\frac{da}{dP}$ | -0.0045 | -0.0049 | -0.0068 | -0.0067 |

| | (Å/GPa) | | | | |
|---|---|---|---|---|---|
| $\frac{dc}{dP}$ (Å/GPa) | -0.0263 | -0.0283 | - | - |
| $\frac{dv}{dP}$ (Å³/GPa) | -0.53 | -0.57 | -0.36 | -0.35 |
| $B'_0$ (GPa) | 6.96 | 4.50 | 6.54 | 4.40 |

### 3.2. High Pressure Resistivity:

Figure 6 shows the temperature dependent resistivity curves for a pressure range of 0 to 2.8 GPa for the temperature range 200-300 K. The exponential increase of the resistivity with the decrease in temperature highlights the semi-metallic nature of the material. Below ~ 210 K, the resistance of the specimen became so high that no current was passed through. Hence, the data is limited to ~ 210 K. As for the pressure dependent ρ(T), the resistivity for each pressure increases monotonically with increasing applied pressure. We do not observe any phase transition up to 2.8 GPa. The monotonic increase in resistivity with applied pressure is consistent with the reported result of NaCoO$_2$ [28]. From Figure 6 (d,e,f), we can observe that the increment in resistivity at the minimum temperature (T$_{Min}$) of 210 K, mid temperature (T$_{Mid}$) of 255 K and maximum temperature (T$_{Max}$) of 300 K, are linear throughout the various applied pressures.

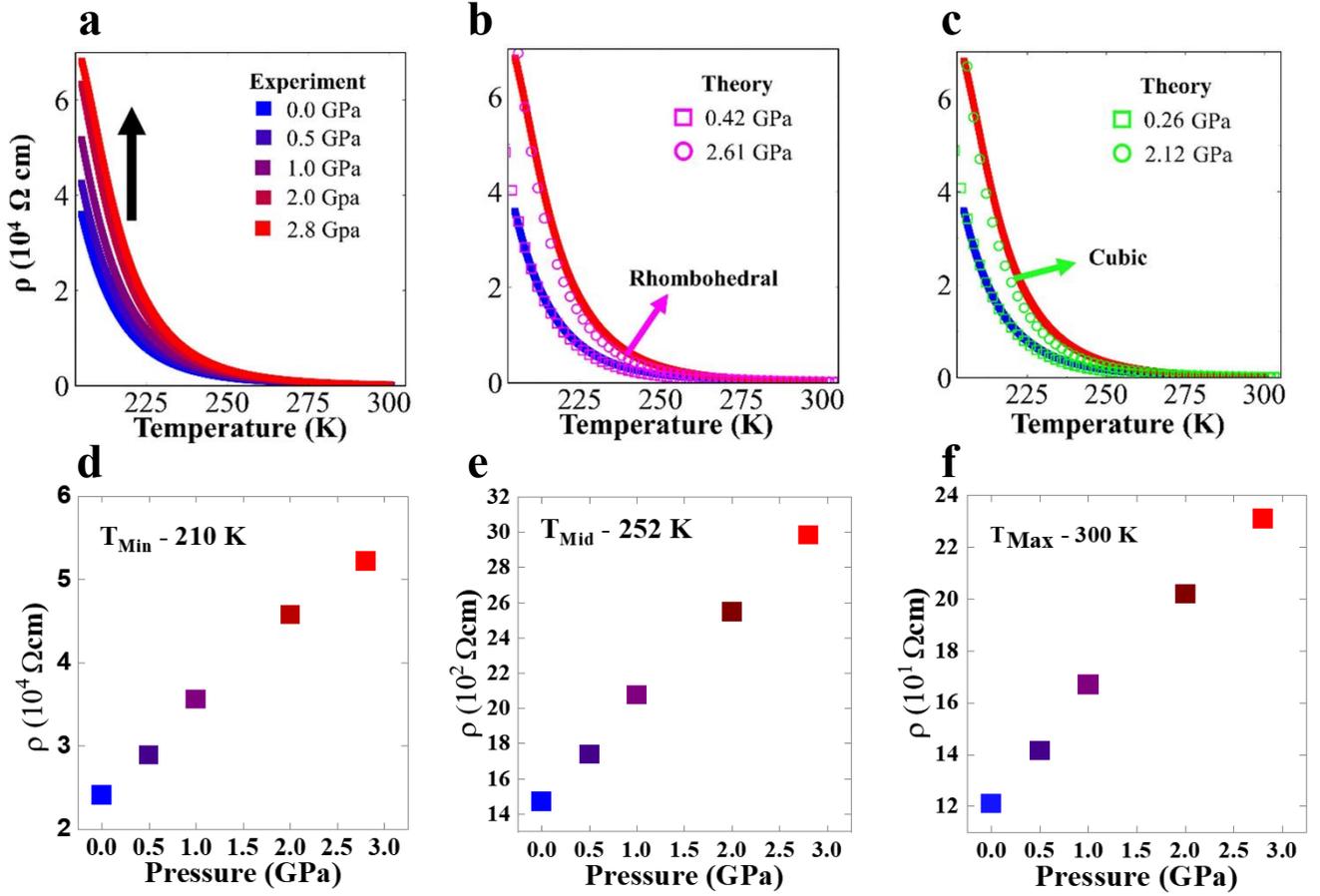

*Fig 6.* (a) ρ as a function of temperature. The interpolated colours from blue to red indicate a variation in applied pressure from 0 to 2.8 GPa, respectively. The vertical black arrow denotes an increase in resistivity near 200 K upon application of external hydrostatic pressure. Calculated ρ overplotted on measured resistivity for (b) rhombohedral (magenta) and (c) cubic (green) phases. Theoretically obtained pressure for experimental structures are shown as legends. (d,e,f) ρ as a function of pressure at fixed temperatures of $T_{Min}$ – 210 K, $T_{Mid}$ – 255 K, $T_{Max}$ – 300 K, respectively. .

In the insulating region, the resistivity phenomenon can be explained by means of a small polaronic hopping (SPH) model expressed as [29,30]

$$\rho = \rho_0 T\, e^{\left(\frac{E_a}{k_B T}\right)}, \qquad (2)$$

where $\rho_0$, $k_B$ and T are the prefactor, the Boltzmann constant and the absolute temperature, respectively. We fit the slope of ln(ρ/T) vs (1/T) plot to extract the activation energy in Figure

7(a,b). The obtained activation energy for the same is plotted in Figure 7(c) which shows that the activation energy of the material increases with increasing pressure.

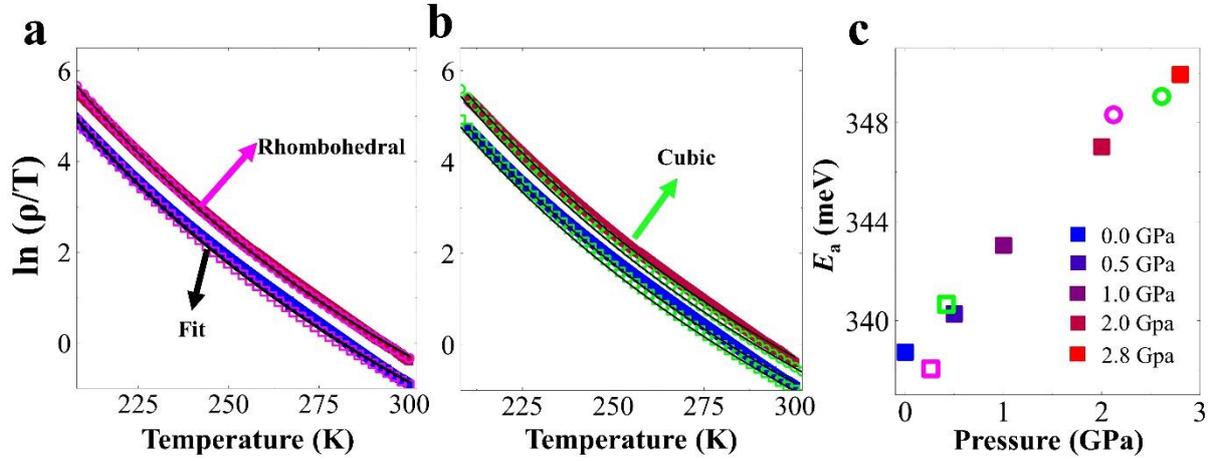

***Fig7.****(a , b) (ln(ρ/T)) vs T to extract activation energies for both rhombohedral (magenta) and cubic (green) phases, respectively for lowest and highest measured pressure (filled symbols) and from simulations (open symbols). Black lines in (a) and (b) denote fits to equation (2). (c) Activation energies obtained from measured and calculated ρ (as described in text).*

From the measured temperature dependent resistivity data for various pressures, the piezo-resistance of the specimen $LiCoO_2$ can be calculated. Piezo-resistivity is the change in resistivity with respect to the applied external pressure and is expressed as [29],

$$PR(T) = \frac{\rho_P(T) - \rho_A(T)}{\rho_A(T)}, \qquad (3)$$

where $\rho_P$ is the resistivity at pressure P and $\rho_A$ is the resistivity at ambient pressure. Figure 8(a) shows the piezo-resistance of the $LiCoO_2$ material at various pressures. Here, we are having positive piezo-resistivity (+ PR) where the resistivity of the material increases as pressure increases. The increase in PR for different pressures for both rhombohedral ($R\bar{3}m$) and cubic ($Fm\bar{3}m$) phases is shown in Fig. 8(b,c). Except below ~ 210 K where resistance data is a little unreliable due to high resistance, simulations reasonably reproduce the trend.

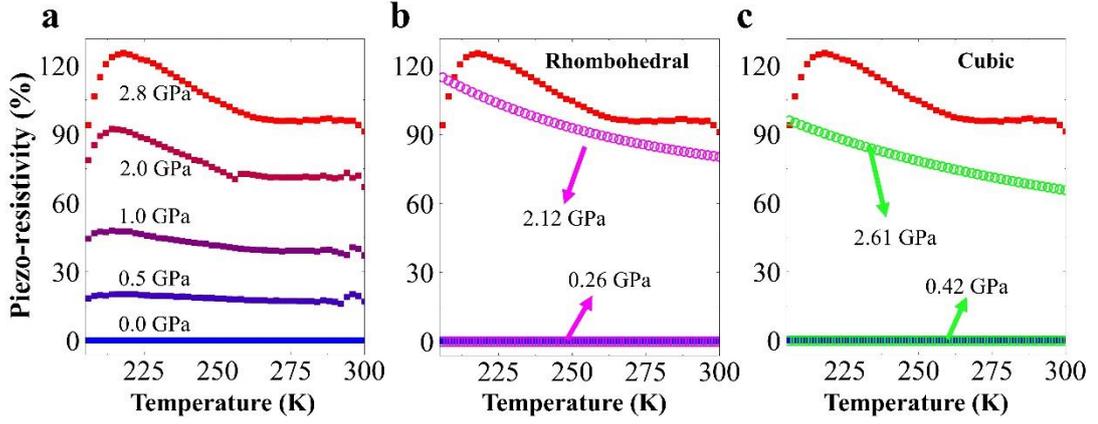

*Fig. 8* (a). *Piezo-resistivity as a function of pressure up to 2.8 GPa. Color scheme follow from Fig. 6(a). (b,c) Comparison of obtained PR percentages from measured (filled markers) and calculated (hollow markers) ρ for rhombohedral (R$\bar{3}$m) and cubic (Fm$\bar{3}$m) phases, respectively. Theoretical pressures are indicated in the inset.*

Further, the electrical transport mechanism in the higher resistance region may be explained by means of the two-dimensional variable range hopping model (2D VRH) or the three-dimensional variable ranging hopping model (3D VRH). Here, we explain the mechanism by adopting the 2D VRH model in order to find the density of states at the Fermi level. The equation for the 2D VRH can be expressed as [31,32],

$$\rho = \rho_0 \cdot \exp\left(\frac{T_0}{T}\right)^{\frac{1}{3}}, \qquad (4)$$

where $T_0 = 16\alpha^3[k_B N(E_F)]^{-1}$, $N(E_F)$ is the density of states at Fermi level, $k_B$ is the Boltzmann constant, and $\alpha$ is the decay constant of the electron wave function (2.22 nm$^{-1}$). Figure 9(a) depicts the above equation fitted on the slope of $\ln(\rho)$ vs $(1/T)^{1/3}$ plot for extracting the density of states at various pressures. The obtained $N(E_F)$ values are shown in Fig. 9(b). The density of states values decreases with pressure indicating the decrement in the charge carriers which in turn increases the resistivity at higher pressures.

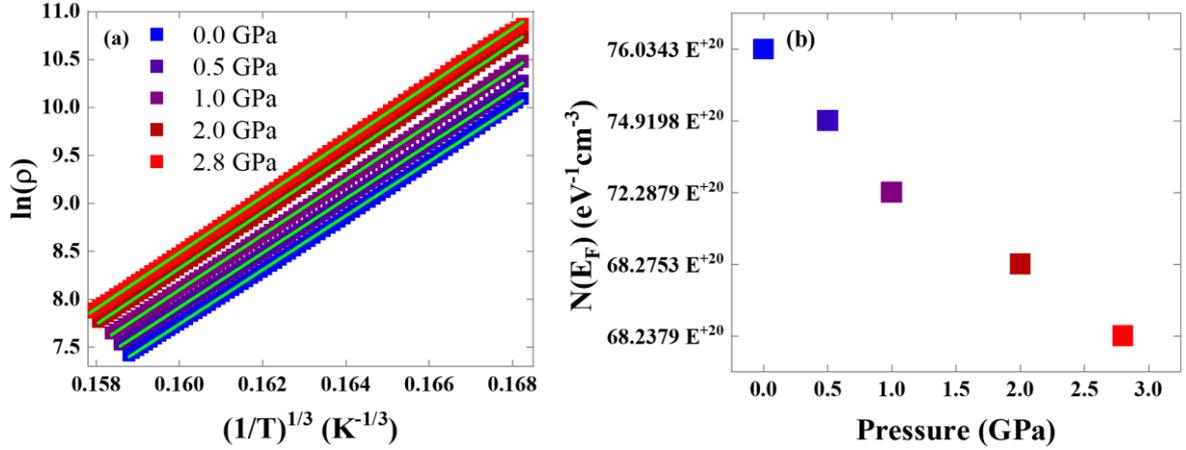

**Figure 9** *(a). 2D VRH fitted over ln(ρ) vs $(1/T)^{1/3}$ for various pressures. (b). The density of states at the Fermi level is extracted from the 2D VRH model.*

## 4. DISCUSSION:

Now we discuss the synthesis and thermodynamic stability of the observed disordered rock-salt structure ($Li_{0.12}Co_{0.88}O$, $Fm\bar{3}m$), and its potential as a high-capacity electrode material. This kind of disordered rock-salt phase has been observed (i) after mechanical alloying of $LiCoO_2$ for a long time (~ 64 hours) [33] (ii) during the solid-state reaction, with $Li_2O_2$ and CoO mixture pressed into pellets in an argon atmosphere and pellet heated in a sealed container (maximum up to $x = 0.2$ in $Li_xCo_{1-x}O$), due to the presence of a phase boundary [34]. In the present study, where the crystals grow out of the melt, the dc magnetization study indicates the presence of the disordered rock-salt phase $Li_{0.12}Co_{0.88}O$ in single crystal samples (Fig. 10), where the peak corresponding to the antiferromagnetic (AFM) transition temperature for $Li_{0.12}Co_{0.88}O$ phase at 251 K has been observed. The $Li_{0.12}Co_{0.88}O$ phase is thermodynamically stable as the magnetization and diffraction measurements were recorded several months after the growth. Here it may be noted that the end member, $CoO_2$, of the solid solution $Li_xCoO_2$, was reported to have $CdI_2$-type trigonal crystal structure (space group $P\bar{3}m1$, lattice parameter $a = 2.82$, $c = 4.29$ Å) [35]. The crystal structure of $CoO_2$ was reported to contain highly repulsive layers of $O^{2-}$. This combined with the fact that cobalt ions in $CoO_2$ are in a 4+ oxidation state makes it thermodynamically unstable and it transforms to CoOOH (space group

$R\bar{3}m$ isostructural with Li$_x$CoO$_2$) in the presence of air at room temperature. In general, the crystal structure of the CoO$_x$ system strongly depends up on the thermodynamic relative stability of the cobalt oxidation states under different synthesis conditions, that is, oxidizing or reducing atmosphere. Recent global potential energy surface calculations [36] have revealed that CoO$_x$ has three thermodynamically stable composition under ambient conditions: (i) wurtzite CoO (P63mc, #186) (ii) rock-salt CoO (Fm$\bar{3}$m, #225), (iii) spinel Co$_3$O$_4$ (Fd$\bar{3}$m, #227), with Co$_3$O$_4$ being the most stable phase. Both wurtzite and rock-salt phases of CoO transform to spinel Co$_3$O$_4$ phase upon annealing at 240 °C under atmospheric air. Upon heating above 320 °C, under an inert, vacuum atmosphere or at high pressure [0.8 − 6.0 GPa at ambient temperature], the wurtzite CoO phase irreversibly transforms to rock-salt CoO, via a reconstructive solid-state phase transition with the breaking of Co−O bonds, having a high barrier. Although calculations show that wurtzite CoO is energetically more stable than rock-salt CoO [37] the possibility of adaptation of structural defects makes rock-salt CoO more stable, as found in experiments where wurtzite phase has only been observed in nanocrystals thus far [38]. The observed disordered rock-salt cubic phase Li$_{0.12}$Co$_{0.88}$O (*Fm$\bar{3}$m*) in the present study possibly also arises due to its capacity to adopt more structural defects (Li and Co share the same site resulting in higher configurational entropy), and this feature further assists in the structural stability of the rhombohedral phase in the presence of large lattice strains.

The phase diagrams of the layer-structured cathode materials for Li-ion batteries, LiMO$_2$ (M = Ni, Co, Mn, and dopants) has recently been investigated using density functional theory calculations. Three phases, the spinel, rock-salt, and layered phases were reported to be thermodynamically stable, except for LiCoO$_2$ where spinel phase was found to be thermodynamically unstable [39]. For Li$_{1.2}$Ni$_{0.2}$Mn$_{0.6}$O$_2$, a phase transformation pathway, i.e., from a layered to a spinel-like (*Fd$\bar{3}$m*) to a disordered rock-salt phase (*Fm$\bar{3}$m*), was described by Zheng *et al.* [40]. This transition was reported to involve a large amount of Li/O release during the phase transition from Li-rich layered to Li-free rock-salt ($Fm\bar{3}m$) structure. The layered rhombohedral structure ($R\bar{3}m$) of LiCoO$_2$, observed in the present study, can be considered as a Li-containing ordered rock-salt derivative, where octahedrally coordinated Co and Li ions perfectly form alternating layers confined to the (111) plane. Historically, disordered rock-salt-type phase Li$_{0.12}$Co$_{0.88}$O (Li$_{1-x}$Co$_x$O-type, cubic), found in the present study, is considered as "electrochemically inactive" and expected to block lithium diffusion pathway and may result in a poor electrochemical performance. However, recent experimental and theoretical research reveals that oxides with the disordered rock salt structure have potential as high-capacity electrode materials [41]. For the stoichiometric Li-excess disordered rock-salt material Li$_{1.211}$Mo$_{0.467}$Cr$_{0.3}$O$_2$ [42], excellent energy density of 660 Wh kg$^{-1}$ at 2.5 V, was reported, which is rarely achieved even in layered Li-transition metal oxides. The observed high energy density was ascribed to percolation of a certain type of active diffusion channels, 0-transition metal channels with no face-sharing transition metal ion, in disordered rock-salt phase, if enough Li excess is available. Moreover, intrinsic disorder on the cation lattice was reported to result in small and isotropic volume changes, as a function of Li concentration, resulting in less mechanical stress and capacity loss. Designing mixed phase materials like the one in this study therefore becomes a viable path that could result in increased structural stability despite increasing lattice strain during Li insertion/de-insertion processes.

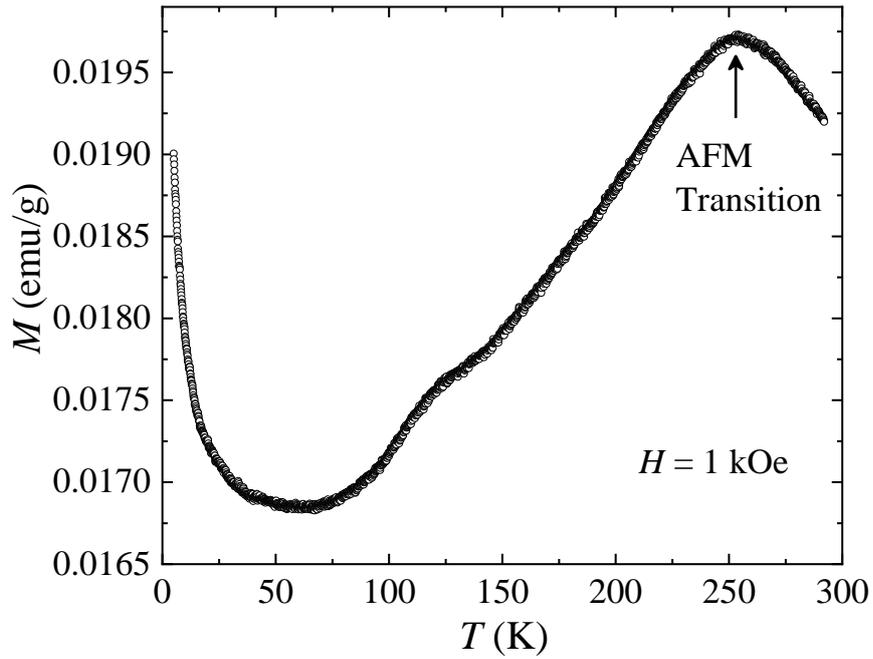

*Figure 10.* *Temperature dependence of magnetization under applied magnetic field of 1000 Oe. Arrow shows AFM transition temperature for $Li_{0.12}Co_{0.88}O$ phase. Reported AFM transition for CoO is 289 K. The observed reduced temperature of 251 K is possibly due to lithium substitution at the cobalt site.*

**Conclusion:**

Detailed high-pressure x-ray diffraction study using synchrotron radiation source and transport measurements are carried out on a $Li_xCoO_2$ single crystal sample, grown under reducing atmosphere (high argon pressure) using the optical floating zone technique, with a view to explore the effect of lattice strain that occurs during lithiation/de-lithiation processes. The co-existence of a significant fraction of $Li_{0.12}Co_{0.88}O$ cubic phase ($Fm\bar{3}m$) along with the $LiCoO_2$ rhombohedral phase ($R\bar{3}m$) is observed in the crushed single crystal sample. The volume fraction of the rhombohedral and cubic phase is 60:40, which remains unchanged up to 10.6 GPa. No structural phase transition has been observed up to 10.6 GPa, indicating substantial structural stability of the two-phase material under large lattice strain. An increase in resistance with a decrease in temperature has revealed the semi-metallic nature of the sample. The derived value of piezo-resistivity increases with applied pressure. Piezo-resistivity at 2.8

GPa is 1.25 % higher than that of the ambient pressure resistivity. A small polaronic hopping model has been adopted in order to explain the conduction mechanism in the insulating region and the activation energy ($E_a$) has been calculated by fitting SPH relation and it is found that $E_a$ increases with increasing applied pressure. Density of states at the Fermi level level, N($E_F$), for various pressures has been extracted by using a 2D variable range hopping model. $N(E_F)$ is found to be decreasing with increasing pressure. The above-discussed results derived from the resistivity confirm that the semi-metallic behaviour was enhanced by the application of pressure. The obtained results can be qualitatively explained via density functional theory (DFT) and thermodynamics modelling. For the rhombohedral phase, the derived value of gradual angle [9.848° and 9.994° at 0 and 10.6 GPa, respectively], which plays a crucial role in layered-to-trigonal phase transformation upon insertion/de-insertion of Li, increases with increasing pressure. The observed increase in the gradual angle, with no change in the lithium contents, further points to the increased stability of the rhombohedral majority phase and also suggests a novel path of applying internal pressure using the dopants at Li and Co sites to synthesize $Li_xCoO_2$, with a view to mitigate the layered-to-trigonal phase transformation in cathode materials for lithium-ion batteries. We also observe that the relative changes in lattice parameters and volume, as a function of pressure, are small for the disordered rock-salt phase ($Fm\bar{3}m$), compared to the layered rhombohedral phase ($R\bar{3}m$). Our results therefore suggest the possibility of the disordered rock-salt phase $Li_{1-x}Co_xO$ acting as a transitional phase during the Li de-insertion/insertion process to and from the layered phase and provides a rationale for the feasibility of designing such mixed phase materials for use as a cathode in lithium-ion batteries.

**References:**

[1]. Y. Chen, Y. Kang, Y. Zhao, L. Wang, J. Liu, Y. Li, Z. Liang, X. He, X. Li, N. Tavajohi, and B. Li, Journal of Energy Chemistry **59** 83–99, (2021)

[2]. E. Hu, Q. Li, X. Wang, F. Meng, J. Liu, J. Zhang, K. Page, W. Xu, L. Gu, R. Xiao, H. Li, X. Huang, L. Chen, W. Yang, X. Yu, and X. Yang, Joule 5, 720, (2021)

[3]. U. Kim, S. Lee, N. Park, S. J. Kim, C. S. Yoon, and Y. Sun, ACS Energy Lett. 7, 11, 3880 (2022).

[4]. A. Van der Ven, M. K. Aydinol, G. Ceder, G. Kresse, and J. Hafner, Phys. Rev. B, 58, 2975 (1998)

[5]. X. Lu, Y. Sun, Z. Jian, X. He, L. Gu, Y. Hu, H. Li, Z. Wang, W. Chen, X. Duan, L. Chen, J. Maier, S. Tsukimot, and Y. Ikuhara, Nano Letters, 12, 6192 (2012)

[6]. C. A. Marianetti, G. Kotliar, and G. Ceder, Nature Materials, 3:9 3, 627 (2004)

[7]. S. Li, Y. Sun, A. Gao, Q. Zhang, X. Lu and X. Lu, Proceedings of the National Academy of Sciences, 119, e2120060119 (2022)

[8]. A. Jain, A. Mohan, and S. M. Yusuf, Journal of Crystal Growth, 536, 125578 (2020)

[9]. X. Chen, H. Lou, Z. Zeng, B. Cheng, X. Zhang, Y. Liu, D. Xu, K. Yang, and Q. Zeng, Matter and Radiation at Extremes, 6, 038402 (2021)

[10]. B. H. Toby and R. B. Von Dreele, Journal of Applied Crystallography. 46, 544–549 (2013)

[11]. K. Murata, K. Yokogawa, H. Yoshino, S. Klotz, P. Munsch, A. Irizawa, M. Nishiyama, K. Iizuka, T. Nanba, T. Okada, Y. Shiraga, and S. Aoyama, Review of Scientific Instruments, 79, 085101 (2008)

[12]. G. Lingannan, B. Joseph, M. Sundaramoorthy, C. N. Kuo, C. S. Lue and S. Arumugam, Journal of Physics: Condensed Matter, 34 245601, (2022)

[13]. G. Kresse and J. Hafner, Physical Review B, 48, 13115 (1993)

[14]. G. Kresse and J. Furthmüller, Physical Review B, 54, 11169 (1996)

[15]. G. Kresse and J. Furthmüller, Computational Materials Science, 6, 15 (1996).

[16]. J. P. Perdew, K. Burke, and M. Ernzerhof, Physical Review Letters, 77, 3865 (1996)

[17]. S. L. Dudarev, G. A. Botton, S. Y. Savrasov, C. J. Humphreys, and A. P. Sutton, Physical Review B, 57, 1505 (1998).

[18]. I. B. Slima, K. Karoui, and A. B. Rhaiem, Ionics (Kiel), 29, 1731 (2023)

[19]. F. Parmigiani, and L. Sangaletti, Journal of Electron Spectroscopy and Related Phenomena, 98–99, 287 (1999)

[20]. A. A. Mostofi, J. R. Yates, G. Pizzi, Y. Lee, I. Souza, D. Vanderbilt, and N. Marzari, Computer Physics Communications, 185, 2309 (2014)

[21]. G. Pizzi, D. Volja, B. Kozinsky, M. Fornari, and N. Marzari, Computer Physics Communications, **185**, 422 (2014)

[22]. F. Khatun, M. A. Gafur, M. S. Ali, M. S. Islam, and M. A. R Sarker, Journal of Scientific Research, **6**, 217 (2014)

[23]. X. Wang, I. Loa, K. Kunc, K. Syassen, and M. Amboage, Phys Rev **72**, 224102 (2005).

[24]. Y. Hu, L. Xiong, X. Liu, H. Zhao, G. Liu, L. Bai, W. Cui, Y. Gong and X. Li, Chinese Physics B **28**, 016402 (2019)

[25]. E. Antolini, Solid State Ion **170**, 159 (2004)

[26]. J. N. Reimers and J. R. Dahn, J Electrochem Soc **139**, 2091 (1992)

[27]. F. Birch, Physical Review **71**, 809 (1947)

[28]. K. Miyoshi, M. Miura, H. Kondo, and J. Takeuchi, Journal of Magnetism and Magnetic Materials 310 901–903, (2007)

[29]. U. Devarajan, S. Singh, S. E. Muthu, G. K. Selvan, P. Sivaprakash, S. R. Barman, and S. Arumugam, Appl Phys Lett **105**, 252401 (2014)

[30]. R. Thiyagarajan, S. E. Muthu, G. K. Selvan, R. Mahendiran, and S. Arumugam, J Alloys Compd **618**, 159 (2015)

[31]. S. Arumugam, C. Saravanan, R. Thiyagarajan, and G. N. Rao, Journal of Magnetism and Magnetic Materials 507, 166775 (2020)

[32]. P. V. Kanjariya, G. D. Jadav, C. Saravanan, L. Govindaraj, S. Arumugam, and J. A. Bhalodia, Journal of Materials Science: Materials in Electronics, 29:8107–8134, (2018)

[33]. M. N. Obrovac, O. Mao, and J. R. Dahn, Solid State Ionics 112, 9 (1998)

[34]. W.D. Johnston, R.R. Heikes, and D. Sestrich, J. Phys. Chem. Solids **7**, 1 (1958)

[35]. G. G. Amatucci, J. M. Tarascon, and L. C. Klein, J. Electrochem. Soc. **143**, 1114 (1996)

[36]. F. Kong, Y. Li, C. Shang, and Z. Liu, J. Phys. Chem. C 2019, **123**, 17539 (2019)

[37]. K. Saritas, J. T. Krogel, and F. A. Reboredo, Phys. Rev. B **98**, 155130 (2018)

[38]. J. F. Liu, Y. He, W. Chen, G. Q. Zhang, Y. W. Zeng, T. Kikegawa, and J. Z. Jiang, J. Phys. Chem. C **111**, 2 (2007)

[39]. Y. Kim, Applied Physics A **126**, 556 (2020)

[40]. J. Zheng, P. Xu, M. Gu, J. Xiao, N. D. Browning, P. Yan, C. Wang, and J. Zhang, Chem. Mater. **27**, 1381(2015)

[41]. R. J. Clément, Z. Lun, and G. Ceder, Energy Environ. Sci., **13**, 345 (2020)

[42]. J. Lee, A. Urban, X. Li, D. Su, G. Hautier, and G. Ceder, Science, **343**, 519 (2014)